\documentstyle[preprint,aps,12pt,epsf]{revtex}

\begin{document}
\draft
\title{ Chaotic scattering in heavy--ion reactions}
\author{ M. Baldo, E.G. Lanza and A. Rapisarda}

\address{  Istituto Nazionale di Fisica Nucleare, Sezione di Catania}
\address{  Dipartimento di Fisica,  Universit\'a di Catania }
\address{  Corso Italia 57, I-95129 Catania, Italy }
\date{june 12th 1993}
\maketitle

\begin{abstract}
We discuss the relevance of chaotic scattering in heavy--ion reactions at
energies around the Coulomb barrier. A model in two and three dimensions
which takes into account rotational degrees of freedom is  discussed both
classically and quantum-mechanically. The typical  chaotic features found
in this description of heavy-ion collisions are connected  with the
anomalous behaviour of several experimental data.
\end{abstract}


\section{Introduction}

The study of classical dynamical chaos has been extended in the last  years
to the case of open systems. It has been found that scattering  variables
have an irregular behaviour as a function of the initial conditions when
the interaction zone is chaotic.  Though  scattering trajectories explore
the real chaotic region only for  a finite time, their behaviour can be so
complicated  that the final observables show strong and unpredictable
fluctuations. These fluctuations are present on all scales of the initial
conditions, revealing an infinite  set of  singularities with a Cantor-like
fractal structure. Singularities  are connected with those trajectories
that remain trapped in the interaction region for very long times. In this
sense the term chaotic has been extended also to scattering situations.
Many examples have been investigated
\cite{eck88,jun87,blu88,gas89,ble90,smi91,jun91,rap91,das92,dro93}  and the
phenomenon is so wide-spread to make one think that it is the rule rather
than the exception. Important consequences of the underlying  classical
chaoticity have been found also in the semiclassical and quantal scattering
counterparts \cite{blu88,gas89,smi91,ohn92,ba92a,ba92b,ba92c,das93}.

In this paper we discuss the occurrence of chaotic scattering  in nuclear
reactions. Investigations on chaotic motion in nuclear  physics started
long ago \cite{boh88} and they have been further stimulated by the recent
progress on dynamical systems \cite{mer88,alh92,blo93,zel93}. Scattering
experiments are one of the main tools to study the nuclear structure,
therefore it is rather important to know their regular or chaotic
character. In particular studies with  heavy ions  cover a wide area of
interest due to the great variety of  nuclear phenomena which can be
investigated \cite{bro81,brom85}.

We consider in particular the reaction between a  spherical and  a deformed
nucleus taking into account rotational degrees of freedom only. This is a
simplified description of the way in which  two nuclei can interact, but it
can be  considered very realistic for many heavy--ion reactions. We show
that  even a few degrees of freedom can produce a very  complicated and
unpredictable motion.

This subject has been already discussed  in several published papers
\cite{rap91,ba92a,ba92b,ba92c},  however, in the following  we review in a
coherent and general  way what has been found including new and more
detailed results. At the same time we try to use a simple and schematic
language in order to explain  even to  the less expert reader  the reason
of  chaoticity  onset, the meaning of it  and the  experimental
implications according to the present understanding.

Unlike other papers in the present focus issue, our  point of  view is more
phenomenological in the sense that we consider  realistic potentials using
the actual units adopted in nuclear physics. On the other hand,  in this
field and in particular in heavy-ion  scattering the knowledge of the
parameters which define the ion-ion potential are  known within  a
10-20$\%$  uncertainty.   Therefore  it would be meaningless to
investigate the peculiarities of the scattering related to the finest
details of the potential. However, adopting the well developed techniques
extensively used for very simple hamiltonians, it is demonstrated that
chaotic scattering in heavy--ion reactions is ubiquitous and does not
depend on these details.  It is shown as well  that chaotic scattering is
not a far out possibility, having real and serious implications which can
be found experimentally. We discuss in particular  in section II and III
the classical dynamics of a  reaction between a spherical and a deformed
nucleus both in two and three dimensions. Actually it is shown that the
former is a particular case of the latter. In section IV the quantal
dynamics is then studied  by means of a coupled channel approach.  The
connection between classical and the quantal treatment  is discussed in
section V.  Realistic quantal calculations are then presented in section
VI. Finally the connection with real experiments is illustrated in section
VII.  A summary is done in section VIII.

\section{Classical scattering }

First of all we introduce a three-dimensional model to describe  the
scattering between a spherical nucleus (1) and a deformed one (2).  Using
polar coordinates, the Hamiltonian depends on 5 degrees of freedoms, i.e.
$r $, $\theta$ and $\phi$ to describe the motion of the spherical
projectile and  $\Theta$, $\Phi$ for the deformed rotor, see fig.1. Thus
the Hamiltonian can be written as
\begin{equation}
H  =  H (r , \theta , \phi , \Theta , \Phi) =  T(r, \theta , \phi) + H_2
(\Theta , \Phi)  + V (r , \theta , \phi , \Theta , \Phi),
\end{equation}
\noindent
where $T$ is the kinetic term, $H_2$ the Hamiltonian of the deformed
nucleus  2 and $V$ the interaction potential. More precisely, $T$ is given
by
\begin{equation}
  T ~ = ~ {{p^2_r}\over{2 m}} ~ + ~ {1\over{2 m r^2}}
   ~ ( ~ p_{\theta}^2 ~ +  ~ {p_{\phi}^2\over{sin^2 {\theta}}} ~ ) ~ ,
\end{equation}
being $m$ the reduced mass and $p_r$, $p_{\theta}$,$p_{\phi}$ the conjugate
momenta of $r$, $\theta$ and $\phi$,  while  $H_2$ is
\begin{equation}
      H_2 ~ = ~  {1\over{2 \Im }} ~
      ( ~ p_{\Theta}^2 ~ + ~ {p_{\Phi}^2\over{sin^2{\Theta}}} ~ )~.
\end{equation}
In this equation $\Im$ indicates the moment of inertia,  while
$p_{\Theta}$ and $p_{\Phi}$ the conjugate momenta of  $\Theta$ and $\Phi$,
respectively.

The ion-ion potential $V$ contains the monopole and quadrupole term  of the
Coulomb interaction plus the nuclear part $U_N$
\begin{equation}
      V = {Z_1 Z_2 e^2 \over r} + {Z_1 Q_0 P_2(cos\xi)\over {2 r^3}}
             + U_N (r,\xi) ~,
\end{equation}
with
\begin{equation}
 cos \xi = cos \Theta cos \theta + sin \Theta sin \theta cos(\Phi - \phi) ~,
\end{equation}
being $\xi $ the angle between the rotor symmetry axis  and the line
joining the centers of the two nuclei. The symbol  $Q_o$ indicates the
intrinsic quadrupole moment, while $P_2$ is the Legendre polynomial of
order 2.  A similar Hamiltonian has already been used \cite{das82} to study
a typical heavy-ion scattering. In our case,  we have chosen  as nuclear
interaction the {\it proximity}  potential \cite{blo77,bro81}. The latter
is extracted taking into account the interaction energy per unit area
between two  curved nuclear surfaces. This  choice has nothing special and
it has been considered only  because this potential is one of the  most
commonly used for deformed nuclei. The  formula of the proximity potential
is
\begin{equation}
         U_N (r, \xi) =
                   4 ~ \pi ~ b~ \gamma ~ \Re ~ \psi (s(\xi)) ~,
\end{equation}
with  the proximity universal function given by
\begin{equation}
{\psi (s(\xi))} = {\Bigg{\lbrace}}
\begin{array}{cc}
{-{1\over2}} ~ (s - 2.54 )^2 ~-~ 0.0852 ~(s - 2.54 )^3
& ~~~~~ if  ~~s \leq 1.2511 \\
{}~~~~-3.437~ exp (- s / 0.75) ~~~~~~~~~~~~~~~~~~~~~~~~~
& ~~~~~~ if ~~ s > 1.2511~. \\
\end{array}
\end{equation}
\noindent
The distance $s$ between the two nuclear surfaces is
\begin{equation}
         s (\xi) ~ = ~ {{r - R_1 - R_2(\xi)} \over b}      ~,
\end{equation}
with   the nuclear radii \cite{bro81}
\begin{equation}
{}~~~~R^o_i = (1.28 A_i ^{1\over 3} - 0.76  + 0.8 A_i ^{-{1\over 3}})
{}~~fm ~~~~~~~ i=1,2
\end{equation}
\begin{equation}
 R_2 ~(\xi)~ = ~ R_2^o ~( ~1 ~+ ~ \alpha_{2 0}~ Y_{2 0}(\xi)~ )~~,
\end{equation}
and the quantity $\Re$ defined as
\begin{equation}
    \Re (\xi) ~ = ~ { R_1 R^o_2 \over { R_1 + R^o_2} }
           \Bigg( ~ 1 ~ - ~ {2 R_1\over{R_1+R_2^o}}
         ~ \alpha_{2 0} ~ Y_{2 0} (\xi) ~\Bigg)                ~.
\end{equation}
In the latter  $\alpha_{20}$ is the deformation parameter and $Y_{20}$ the
spherical  harmonic of order 2.  The quantity $~\gamma~$  is the so-called
surface tension and it is given by  \cite{bro81}
\begin{equation}
     {\gamma } = 0.95 \Bigg\lbrack 1. -1.8~
\Bigg({{N_1-Z_1}\over A_1} \Bigg)
\Bigg({{N_2-Z_2}\over A_2} \Bigg) \Bigg\rbrack
{}~~~MeV~fm^{-2} ~,
\end{equation}
being $N_i$  and $A_i$ the neutron and atomic numbers of the two nuclei,
while $~b~$ is the diffuseness parameter which is equal to 1 fm. In the
following the units which are used are those  commonly adopted in nuclear
physics, that is $~fm~$ for distances and $~MeV~$ for energies. The actual
value of $\hbar$   has been  considered, more precisely it has been  used
$\hbar~c~ =~ 197.329 ~ MeV~ fm$.

It should be noted that eq.(8) represents the coupling between the relative
motion and the internal (rotational) degrees of freedom.  This coupling,
breaking the central symmetry  of the potential, is the one  responsible
of the onset of chaotic scattering as discussed later.

Solving the equations of motion for  the Hamiltonian (1) one can follow in
time the evolution of the  system. However these equations are very general
and complicated, thus  in order to show in a clear and simple way the
typical features of chaotic motion, let us consider for the moment  the
scattering occurring on the  x-y plane. In  this case we have only 3
degrees of freedom, i.e. $r~$, $~\xi$ and $~\phi$, and the Hamiltonian
reduces to
\begin{equation}
{\cal H}  =  {p^2_r\over{2 m}} + {I^2 \over {2\Im}} +
{\ell^2\over{2 m r^2}} + V(r, \xi )  ~,
\end{equation}
where  $\xi = \Phi - \phi$, and $ \ell=L-I$ is the orbital angular momentum
with $L$ and $I$ the total angular momentum  and the spin of the rotor,
respectively.

The equations of  motion corresponding to the Hamiltonian ${\cal H}$ are
therefore
\begin{equation}
\begin{array} {ccccc}
  {\dot r }  & ~ = ~& {{p_r\over m}}   & ~ & ~ \\
  {\dot p_r}  & ~ = ~&  {(p_\phi - p_\xi)^2 \over 2mr^3} -
  {\partial V(r,\xi) \over \partial r} & ~ & ~  \\
  {\dot \xi } &~ = ~ & {I\over \Im} - {(L-I)\over mr^2}
  &  ~  &   ~  \\
   {\dot I}   &~ = ~ & {\dot p_\xi}   &
  ~ = ~& - {\partial V(r, \xi ) \over \partial \xi }\\
  {\dot L}   &~ = ~& {\dot p_\phi} & ~ = ~&~ 0 ~\\
\end{array}
\end{equation}

\noindent
Our Hamiltonian has two constants of motion, namely the total energy $E$
and the total angular momentum $L$, as it can be seen from the last of
eqs.(14).  Neglecting the $\xi$-dependence of the full ion-ion potential,
the Hamiltonian is separable and thus integrable, because the internal
angular momentum $I$ and the orbital one $\ell$ are conserved separately.
However the $\xi$-dependence of the ion-ion potential introduces a
symmetry-breaking term leading  to the conservation of $L$ only and
generating  the onset of chaos. In reality  the scattering problem is
integrable asymptotically. It is  the chaoticity of the interaction zone
which makes the scattering become chaotic.  The set  of unstable phase
space trajectories which are confined in the interaction region defines the
so-called {\it repeller}. The latter has an unstable manifold which
extends to asymptotic distances, thus  scattering trajectories are trapped
for long but finite times inside the phase space region. The erratic,
though deterministic, motion of these trapped trajectories, which are those
that come closest  to the repeller, cause the unpredictability of the
final scattering observables on  all scales.

In the following we solve numerically eqs. (14) for the planar case  in
order to illustrate the regular or chaotic character of the nuclear
scattering.  As a first example we take into account the reaction between
the $^{28}Si$ nucleus  considered spherical  and the deformed $^{24}Mg$.
The values adopted for the deformation parameter  $\alpha_{20}$ and  the
quadrupole moment $Q_o$ taken from ref. \cite{bomo,bro78} are reported in
table 1.   The potential $V(r,\xi)$ is shown in fig.2 as a function of $r$.
The dependence  on the angle, for the cases $~\xi=0^o$ and $90^o~$,  is
illustrated for three initial orbital angular momenta $~\ell=15,35,45
{}~\hbar~$ . The change of the orientation angle $\xi$ from $90^o$ to $0^o$
lowers  the height of the barrier and shifts the position of the  minimum
towards larger radii. Increasing $\ell$ the attractive pocket tends to
disappear due to the enhancement of the centrifugal barrier. One should
note that this is only a static picture. Actually, as the nuclei approach
each other,   due to the coupling between the relative motion and the
internal degrees of freedom, the potential  oscillates according  to the
variation of the orbital angular momentum $\ell$ and the angle $\xi$. In
this sense the potential under  investigation is more complicated than the
one of the 3-disks problem \cite{gas89} and  at the same time very
realistic. In fact potentials  of the type  considered here are commonly
used - with different units -  in atomic and molecular physics.

In fig.3 we show, for a fixed total angular momentum $L$, the final  value
of the scattering angle $\phi_f$ as a function of the initial rotor
orientation $\Phi_i$. The initial value of  $\phi_i$ was always set equal
to zero (then $\Phi_i=\xi_i$ ), while the  rotor was considered always at
rest $I_i=0$ (then $\ell_i=L$).  This choice has been kept through all the
calculations presented here. The different trajectories were obtained
varying  the initial angle $\Phi_i=\xi_i$ from $0^o$ to $180^o$ and taking
into account 1000 trajectories  for each of the three different values of
energy shown in fig.3. Below the Coulomb barrier - $V_{B}\sim 26.5 MeV$ -
(bottom panel) we have a regular  and smooth behaviour, while  wild
fluctuations show up as soon as the energy is increased (middle panel).
These fluctuations tend to vanish and give again a regular motion with only
a few  singularities as the energy is further increased. In fig. 4 we show
the deflection function, i.e. the final scattering angle as a  function of
the total angular momentum. In this case the orientation angle was  fixed
to the initial value $\Phi_i=\xi_i=0^o$ while the total angular momentum
was varied. Part (a) of the figure shows strong oscillations of the
deflection function  in between regular regions. Two successive blow-ups
(b) and (c) illustrate  the persistence of these fluctuations at smaller
scales with a very similar structure.  This is the typical manifestation of
chaos in scattering processes   \cite{eck88,jun87,blu88,gas89,ble90,smi91}:
an infinity of singularities having a fractal pattern shows up.  Figures 3
and 4 prove that for the heavy--ion system $^{28}Si + {^{24}Mg}$ the
scattering is chaotic just above the Coulomb barrier.  Only above the
barrier scattering trajectories can probe the chaoticity  of the internal
zone. In order to illustrate the dynamics inside the  pocket we can study
the evolution of bound phase space trajectories.  In fig.5 we display a
Poincar\'e surface of section for ten confined   orbits changing the
deformation parameter $\alpha$. While for $\alpha=0.1 \alpha_{20}$ the
motion is completely regular, when $\alpha$ is equal to the value
$\alpha_{20}$ corresponding to the  deformed nucleus $^{24}Mg$ one has a
real chaotic dynamics.  The first KAM tori start to break around
$\alpha=0.15 \alpha_{20}$. A magnification of the middle panel of fig.5 is
displayed in fig.6 where 90 trajectories are considered.

The system  $^{28}Si + {^{24}Mg}$  has no special characteristics and in
fact we will show in the following that irregular scattering is rather
typical for  light heavy-ions, i.e. nuclei whose atomic mass number A lies
in the range between A=4 and A=60.  In figs.7-9,  the final rotor spin $I$
( in units of the maximum spin $I_{max}={E\over{ 2 \Im}}$ ) ,  the final
scattering angle  $\phi_f$ and the reaction time $T_f$ are shown as a
function of the initial rotor orientation for the systems $~^{4}He +
{^{24}Mg}~$,$~^{12}C + {^{24}Mg}~,$ $~^{86}Kr + {^{24}Mg}~$.  The reaction
time is defined as the time the system takes to go from an  initial
asymptotic distance to the final asymptotic one passing through the
interaction region.  Both initial and final distances are set equal to r=18
fm.  A cut-off time equal to T=$10^4$ fm/c is chosen for those trajectories
which remain trapped inside the nuclear pocket. The same fluctuations
found for $^{28}Si + {^{24}Mg}$   and  characterizing chaotic scattering
are evident for these systems as well.  Different values of energy and
angular momentum are considered to show  that chaotic features are not
present only in a limited region.

On the contrary for the system $~^{86}Kr + {^{152}Sm}~$ only a regular
motion  of the kind shown in fig.10 is found by changing both E and L.
This different behaviour has two main reasons. First, as the atomic number
of the nuclei increases the enhanced Coulomb repulsion reduces the
attractive nuclear pocket. Second, the greater are both the mass of the
nuclei and the moment of inertia the slower is the  motion of the barrier.
The relative motion becomes   faster than the one of the internal  degrees
of freedom,  whose slow variation is not able to  raise the barrier and
trap the spherical nucleus. Therefore, in order to have chaotic motion the
two characteristic time scales should be comparable.

Chaotic scattering is not peculiar of the  simple 2-dimensional  model.  In
fact  taking into account the more general 3-dimensional Hamiltonian (1)
the possibility for the scattering to be chaotic can even increase. This is
shown in fig.11, where the results obtained solving the equations of motion
corresponding to the  Hamiltonian (1)  are displayed for the reaction
$^{12}C + {^{24}Mg}$. In this case the symmetry axis of the deformed target
does not lie completely  on the x-y plane, being $\Theta_i=89^o$. The
reaction does not occur on the plane and, in contrast to the previous
cases, the  angle $\theta$ is not constant as a function of time. In
particular  in correspondence of the irregular  regions the  final
$\theta$-values can be very different from the initial one, see fig.11.
In general, when solving the 3D equations,   if the planar symmetry is
initially assumed it is also maintained throughout the reaction.  In this
sense the 2D scattering is a particular case of the more general 3D model.
However, if a small initial  symmetry-breaking occurs - as in the case
shown in fig.11 -  then  the system explores the complete 10D phase space.
In the case shown in fig.11 the number of trapped trajectories is greater
than in the planar case  of fig.8. This is not true in general.

All these features will be studied more quantitatively in the next section.

\section{Quantifying chaos}

After this qualitative introduction which illustrates the ubiquity of chaos
in light heavy--ions, in this section we take into account the system
$^{28}Si + {^{24}Mg}$ as a typical example and we investigate chaotic
scattering in a quantitative  and detailed way. Possible differences
between the 2D and the 3D case are also investigated and discussed.

In fig.12  the final scattering angle $\phi_f$ as a function of the initial
rotor orientation $\Phi_i$ is shown for four different small intervals, of
initial conditions, i.e. $\Delta \Phi_i=10^o,1^o,0.1^o,0.01^o$. The planar
scattering - panels (a)-(d) - are shown in comparison with the three
dimensional  case - panels (e)-(h).  In this example the total angular
momentum is zero and the total energy is 25 MeV. No clear distinction is
qualitatively evident  in the two cases, nor   the successive blow-ups do
reveal any  deeper  difference in the underlying structure.

In order to study possible quantitative differences, let us calculate the
fractal dimension of the repeller.   From the final scattering angle
reported in fig.12 one can construct the classical cross section counting
the number of final angles which fall inside bins of finite size. This
cross  section as explained in ref. \cite{jun91} presents very many peaks
in correspondence of the (rainbow) singularities which exist around the
extrema of the small regular regions.  It can be shown \cite{jun91} that
the fractal dimension of the rainbows distribution is equal to the fractal
dimension of the repeller. To calculate this fractal dimension the sandbox
method has been used as suggested in ref.\cite{jun91,tel89}.  We use sets
of $10^4$ trajectories to evaluate the classical cross section  $P(\phi_f)$
for the final scattering  angle. As a typical example we show in fig.13 (a)
the one obtained for the planar scattering of fig. 12(c).  The sandbox
method consists in counting the number of angles $N$ entering into circles
of diameter $R$, using as centers the most  pronounced peaks. The average
of $~1/N(R)~$  over the several centers adopted should scale like $R^{-D}$,
where D is the fractal dimension. This method has been proved to be more
efficient  than the box-counting technique \cite{tel89}, but the result
gives often an  estimate which is slightly larger than the true fractal
dimension \cite{jun91}. In fig.13 (b)  we plot the points obtained with the
sandbox method for the cross section shown in fig.13 (a).   In this case
the number of rainbows used as centers is 107 and the bin size  for the
angle is $\Delta \phi_f=0.1^o$. The points follow a straight line over
almost 4 decades showing small deviations only for large R.  A fit of the
slope gives a fractal dimension D=0.73.  To check the accuracy of this
value we have also calculated the uncertainty dimension according  to
ref.\cite{lau91}. This should be less or equal than the real fractal
dimension, and therefore should give us a minimum value. The method of ref.
\cite{lau91} in this case consists in calculating for a fixed uncertainty
$\epsilon$ the quantity $\Delta T (x_o,\epsilon)=|T(x_0)-T(x_0 + \epsilon
)|$,  where T is  the reaction time corresponding to an initial condition
$x_0$ randomly chosen. If $\Delta T (x_o,\epsilon) $ is  greater than a
fixed small quantity ( we used 50 fm/c in our case,  but the actual value
is not important) then one  says that $x_0$ is $\epsilon$ uncertain. The
probability $f(\epsilon)$ to obtain  an initial value which is  $\epsilon$
uncertain - approximated by the ratio between the number of  random calls
which are uncertain and the total number of calls ($10^3$  in our case) -
should scale like $\epsilon^{1-D}$.  Therefore plotting $-Log_{10}
f(\epsilon)/ \epsilon$  versus $Log_{10}\epsilon$,  one gets a line whose
slope is $D$. In order to distinguish between the two  methods we indicate
the corresponding values with  a superscript $u$ for {\it uncertainty} and
$s$ for {\it sandbox}. An example of the determination of  $D^u$ is given
in fig.14 for the same case shown in fig.13.  Also in this case the points
follow very nicely a straight line over almost 4 decades. The value of the
uncertainty dimension is in this case $D^u$=0.84.

The analysis described above has been done   both for the 2D and the 3D
scattering for those small intervals of initial conditions considered in
fig.12.  In all the cases considered a behaviour  similar to that shown in
figs.13 and 14  is found. The results are summarized in tables 2 and 3. In
general the sandbox dimension is not  equal to the uncertainty dimension
and their  relative difference is not always the same.  This is probably
due to the numerical accuracy, being these kind of calculations very
delicate. In order to obtain  a value of fractal dimension closer to the
real one we have taken the average between the two estimates $D^u$ and
$D^s$.  This value is indicated as $\overline{D}$ in the above mentioned
tables.   If the scattering is fully chaotic - hyperbolic scattering - an
exponential law is expected \cite{gas89} for the reaction time
distribution probability $P(T)$.  In this case the repeller is
characterized by   the escape rate  $\Gamma$ defined by the formula
$P(T)\sim exp(-\Gamma T)$.   In tables 2 and 3 it is also reported the
values of $\Gamma$ extracted  by the exponential fits of the time
probability distributions displayed in  fig.15. In this figure it is shown
the logarithm of the collision time probability  as a function of collision
time for the small intervals of initial conditions  into examination. For
very long reaction times an exponential  law fits very well the
distributions, though  in some cases strong peaks are evident.

When the scattering is nonhyperbolic, i.e. there is a coexistence of KAM
surfaces  and chaotic regions \cite{gas89,lau91,ott92}, a power law is
predicted $P(T)\sim T^{-z}$. It is a fact that all  the time probability
distributions of fig.15  can be fitted as well adopting   a power law. The
values of $z$ obtained are reported  in tables 2 and 3. The comparison
between the two kinds of fit are shown in  fig.16 for  a typical case.
This fact supports the conjecture that  our system shows both hyperbolic
and nonhyperbolic features.  Another indication going in the same direction
is the trend of the fractal dimension.  As claimed in   \cite{lau91}, for
hyperbolic scattering  the fractal dimension should be the same when going
to smaller and smaller intervals. This  does  not seem to be always true
for the values obtained, see tables 2 and 3. A clearer answer to this
question could be probably found by looking for invariant tori or surfaces
by means of Poincar\'e sections. A detailed  study along this line is left
for the future.

{}From $\overline{D}$ and $\Gamma$ one can calculate the average  Lyapunov
exponent $\overline{\lambda}$ according to the formula  $\overline{\lambda}
= \Gamma /(1-\overline{D})$  valid for chaotic repellers  \cite{tel87}.
These average exponents are also listed in the tables 2 and 3. Their
variations, though limited in a small range and reflecting  the
fluctuations of  $\overline{D}$, indicate that   the degree of chaoticity
can  depend on the specific intervals taken into account.  Since the
scattering is not exactly hyperbolic,  the Lyapunov exponents thus obtained
should be compared with those  calculated by means of the more standard
formula
\begin{equation}
 \lambda_{\infty}\, = \, \lim_{d_0\to0}\,
\lim_{T\to\infty}\, \lambda(T)~,
\end{equation}
with $\lambda(T)$ defined by
\begin{equation}
                  \lambda (T)\, =\, {\log (d(t)/d_0) \over T} ~.
\end{equation}
The quantities $d(t)$ and $d_0$ are the distance between  two close
trajectories at time $T$ and at time $T=0$ respectively. In fig.17  it is
shown the behaviour of $\lambda(T)$ as a function of time for ten
trajectories chosen randomly in the smallest interval of initial conditions
considered, i.e. $\Delta \Phi_i~ =~ 0.01^o$. The calculations were done
both in the 2D, part (a), and in the 3D, part (b),  case. In the figure,
$\lambda(T)$ shows a clear tendency to have an asymptotic limit
$\lambda_{\infty}$.  The average  $\overline{\lambda}_{\infty} $ over the
trajectories considered is reported in the two cases.  These Lyapunov
exponents are very close to those  $\overline{\lambda}$ values obtained
previously. However it should be noted that the values of
$\overline{\lambda}$ listed in the  tables are probably a better estimate
of the average degree of chaoticity in the specific interval. In fact both
$\overline{D}$  and $\Gamma$ have been calculated by sampling $10^4$
trajectories \cite{lyapnote}.

In conclusion a detailed investigation of classical heavy--ion scattering
reveals a dynamics  which seems to have both hyperbolic and nonhyperbolic
features and,  in contrast to very schematic models,  a greater complexity
of the flow in  phase space which is typical of more  realistic potentials
\cite{dro93}.

\section{Quantum scattering }

In the following we discuss the quantal analog of chaotic scattering
between a spherical and a deformed nucleus.  There is a general agreement
to call {\it quantum chaos} the quantum counterpart of those classical
dynamical systems which show chaotic motion. It has been found that quantum
chaos presents a behaviour which is different from the quantum counterpart
of integrable systems \cite{gut90,jen92,cas91}. The quantum analog of
classical scattering is given by the solution of the Schr\"odinger
equation. In nuclear physics one usually introduces the deformation degrees
of freedom as excited quantum states belonging to rotational bands. These
states are coupled among each other and these  couplings influence the
quantum-mechanical evolution of the reaction. This method is often referred
to as the coupled-channels approach.

We assume for simplicity and only for the moment that the reaction occurs
on a plane.  More specifically, considering planar geometry the
Schr\"odinger equation can be written as
\begin{equation}
{ \bigg[ {-\hbar^2
\Delta^{(2)} \over {2 m}}  + {\ell^2 \hbar^2 \over {2 m r^2}} +  {
{I^2\hbar^2 } \over {2 \Im} } + V(r,\xi) - E \bigg] \, \Psi (r,\xi,\phi) =
0 \,,}
\end{equation}  \noindent
where as before $m$ is the reduced mass, $\xi$ is the rotor orientation
angle, while  $\Delta^{(2)}$ indicates the Laplacian in two dimensions.

{}From eq.(17), after elimination of the angular dependence of the wave
functions \cite{ba92b},  one gets the radial coupled--channels equations
\begin{equation}
{\bigg[ {d^2\over dr^2}  + { 1\over r} {d\over dr} - {\ell^2\over r^2} +
k^2_{L-\ell}(r) \bigg] \psi^L_{\ell}(r) - {2m\over {\hbar^2}} \,
\sum_{\ell'\neq \ell} \,
\,V_{\ell'-\ell} (r) \,\,\psi^L_{\ell'} (r) =0}
\end{equation}
with
\begin{equation}
  k^2_{L-\ell}(r)\, = \, {2m\over {\hbar^2} } (E - E_{rot} - V_o(r))
\end{equation}

\begin{equation}
  E_{rot}\,  = \, {I^2 \hbar^2 \over {2 \Im}}
\end{equation}

\begin{equation}
  V_{\ell'-\ell} (r) \, = \, {1\over {2 \pi}}
     \int_{-\pi}^{\pi} e^{i (\ell'-\ell)
    \xi} \,  V(r,\xi) \, d \xi
\end{equation}

\begin{equation}
  V(r, \xi) \, = \, V_{coul} + V_{nucl} \,\,.
\end{equation}
\noindent
The moment of inertia $\Im$  as well as   the ion--ion potential $V(r,\xi)$
are the same considered in the classical case. To simplify the calculation
the Coulomb tail of the interaction has been taken away. This hardly
influences the scattering which is characterized mainly by the nuclear
interaction. In eq.(21) the coupling is taken to all orders and only
between the nearest neighbours. For more details cfr. refs.
\cite{ba92b,ba92c}.  In principle the summation in eqs.(18) should include
an infinite number of channels, in practice one considers only the most
important $N$ channels. Each of the $N$ eqs.(18) was integrated numerically
from the most internal turning point up to  an asymptotic distance. There
the wave function is the free outgoing solution -  a Hankel function in
this case - times a coefficient which represents the  S--matrix element
\quad $S^L_{I,I'}$ \quad for the particular entrance and  outgoing channels
( $I$ and $I'$ respectively) considered. Solving then the  system of
equations thus obtained for different channels, one can construct the
complete  S--matrix.

Let us consider again as a typical example the reaction $^{28}Si +
^{24}Mg$. Fig.18 shows \quad  the elastic ($I'=0$) and two inelastic
($I'=\pm 2~ \hbar$) transition probabilities  \quad $\big\vert
S^L_{I,I'}(E) \big\vert ^2$, \quad calculated  at total angular momentum
$L=15 \, \hbar$, as a function of incident energy.  For an initial spin
$I=0 \,\hbar$, 11 final channels were considered,  $I=0,\pm 2, \pm 4, \pm
6, \pm 8, \pm 10 \, \hbar$.  An energy step of 20 KeV was adopted for the
calculations.

The S-matrix elements show rapid oscillations as a function of energy with
a width ranging between 50 and 400 KeV, implying   the occurrence of
long-living intermediate states of the  dinuclear system. Resonances
exhibit larger widths,  until their  complete disappearance, as the energy
is increased.  A reduced energy step does not reveal any further
structures. In ref. \cite{ba92a,ba92b}  it was shown that fluctuations
manifest themselves  only in the region of energy and angular momentum
where classical chaos shows up, that is around the potential barrier.  In
this sense we can say that this irregular behaviour is the manifestation of
quantum chaos.

Another indication which gives support to this claim is the fact that the
appearance of sharp and grouped structures depends in a sensitive way on
the strength of the coupling term. In ref. \cite{ba92c} it was demonstrated
that, by decreasing the strength,  fluctuations rarefy and then they
disappear completely. On the contrary, an increase of the coupling produces
an enlargement of the energy  region where fluctuations are present.  In
the example of fig.18 the coupling term adopted comes out of eqs.(16-20)
taking into account the same potential used in the classical chaotic
dynamics.

A quantitative study of the fluctuations  shown in fig.18 can be obtained
by means of autocorrelation  function analysis \cite{blu88,smi91}.  In our
case the autocorrelation function for the  S-matrix elements can be defined
as
\begin{equation}
{  C_{I,I'}(\epsilon)\, =\, <\,S^{\bf *}_{I,I'}(E) \, S_{I,I'} (E+\epsilon) \,>
}   ~,
\end{equation}
where $<\,>$ denotes the average over an appropriate energy interval
$\Delta E$.

For the most populated exit channels, $I'=0,\pm 2 \,\hbar$, the  modulus
square of the autocorrelation function obtained from the  analysis of the
fluctuations is  reported in fig.19 (open squares).  An energy interval
$\Delta E$ =4 MeV is adopted to perform the  averages and the smooth
S-matrix part on this interval is subtracted. The autocorrelation functions
present a lorentzian-like behaviour which reminds of the Ericson's
fluctuations theory  \cite{eri60,bri63,eri63,eri66} developed in the 60s
for compound nucleus reactions.  The connection is discussed in the next
sections. In the figure, lorentzian fits (solid curve)  allow to extract a
coherence length $\Gamma_{quan}$, also shown, that tells us the energy
interval  in which the S-matrix is correlated with itself. The value of
$\Gamma_{quan}$ varies from 80 to 200 KeV. In the elastic channel, $I'=0
\,\hbar$, although some deviations from a lorentzian shape are evident, one
extracts a coherence length which is almost a factor of 2 bigger than the
one corresponding to the inelastic channels.

As discussed also in section VII it is  not clear to what extent one can
link the properties of the S-matrix  fluctuations found here to the
well-known features of Random Matrix Theory and Ericson's theory as was
done in ref. \cite{blu88,smi91,lew91}. Due to the limited number of
resonances one is not able to study their distribution. However, it is
usually claimed  that RMT predictions  apply only to the universal aspect
of chaos,  while in our case  specific  characteristics of the system are
not neglected and can interfere with the universal ones.

\section{semiclassical considerations}

In order to link quantitatively  the classical dynamics with the quantal
counterpart one should perform a semiclassical  analysis.  In this section
we perform this investigation even though in our case  we are not allowed
to use a priori the  semiclassical approximation.

It has been demonstrated \cite{blu88,smi91} that,  in the semiclassical
limit,  one can calculate a semiclassical coherence length $\Gamma_{cl}$
considering the classical distributions of delay time. The latter is
defined as the time the system spends in the interaction region.

In fig.20 we show, for the system $^{28}Si + {^{24}Mg}$  the  delay time
distributions calculated in the classical case for  $10^4$  trajectories.
The energies considered are E=28,28.5,29 MeV and the total angular momentum
is L=15 $\hbar$.  The collision times corresponding to the trivial
reflection from the outer barrier (600-1200 fm/c) were taken away in order
to obtain the real delay time distribution.

A  Fourier transform of $P(T)$  allows to calculate the autocorrelation
function, which is given by
\begin{equation}
C(\epsilon)= \Big \vert \int P(T) e^{i \epsilon T} dT \Big\vert^2 ~~.
\end{equation}
\noindent
According to eq.(24), if the time distribution is exponential \quad P(T)
$\sim $ exp(-$\Gamma_{cl}$ T) \quad then the autocorrelation function is  a
lorentzian \quad $C(\epsilon)= C(0)/ (1 + \epsilon^2/\Gamma^2)$ \quad with
width $\Gamma$.

In fig.20 (c,d,e) the autocorrelation functions correspondent to the
classical delay time distributions of fig.20 (a,b,c) are shown as open
squares. A lorentzian fit is also reported as full curve. Small deviations
from the  lorentzian curve are present as oscillations in fig.20 (e) and
can be attributed to the fact that  $P(T)$  presents  fine structures
superimposed on an exponential background. This is in general true as it
will be discussed later.

Considering again the lorentzian fits one can extract the widths
$\Gamma_{cl}$ =100,130,150 KeV corresponding to the energies  E=28,28.5,29
MeV.  The widths reported above are slightly smaller than those in ref.
\cite {sis92a} due to the elimination of the first peaks in the reaction
time distributions. If one takes an average of the semiclassical coherence
length over the same interval of incident energy  chosen in the quantum
case, the value $\overline{\Gamma}_{cl}$ =250 KeV is obtained.

Comparing then the quantal and the semiclassical values, it turns out that,
as in ref.\cite{blu88,smi91} $\Gamma_{cl}$ and $\Gamma_{quan}$ are equal
within a factor of two.  This nice agreement links quantitatively classical
chaos with its quantum counterpart. At the same time it represents a
surprising result since in  our problem only the lowest channels are
excited and therefore in principle the semiclassical approximation should
not work for the rotational degrees of freedom.

Autocorrelation function deviations from a lorentzian curve  are present
both in the classical and quantal case. In order to understand these
deviations let us  consider a simulated delay time distribution which
mimics the peaks present in fig.20 (b). This is obtained by summing   an
exponential term plus a gaussian peak
\begin{equation}
       F(T) = A ~ {e^{-\Gamma T}} ~ + ~ B ~
  {e^{{-(T-T_o)\over{2 \sigma^2}}}}
\label{ft}
\end{equation}
with $A$ and $B$ constant quantities, see fig.21 (a). The corresponding
Fourier transform $f(\epsilon)$ is
\begin{equation}
       f(\epsilon) = ~A~  {{1\over{\Gamma - i \epsilon}}} +
 ~ B ~  {  e^{i \epsilon T_o }
  e^{ \epsilon^2 \sigma^2 \over 2 }}  ~.
\label{fe}
\end{equation}
Then the autocorrelation function $C(\epsilon)$ is given by the modulus
square of $f(\epsilon)$. The phase in front of the  second term of eq. (25)
gives rise to oscillations in  $C(\epsilon)$ whose magnitude depends on the
height of the gaussian peak with respect to the exponential background.
This is illustrated in fig.21 (c) where the analytical $C(\epsilon)$ (full
curve) is plotted in comparison with the numerical evaluation of the
Fourier transform, open squares. Adding other gaussian peaks  of the kind
$C~ {e^{{-(T-T_i)\over{2 \sigma^2}}}}$ to eq. (25) one should add other
terms  of the kind $  C ~  {e^{i \epsilon T_i } e^{ \epsilon^2 \sigma^2
\over 2 }}  $  to eq. (26). Thus other phases enters into $f(\epsilon)$ and
the behaviour of $C(\epsilon)$ becomes even more complicated. This is
illustrated in fig.21 (b,d), where two gaussians peaks are considered. In
general the analytical formula is very well reproduced by the numerical
calculation of  $C(\epsilon)$, open squares. The dashed lorentzian curves
shown in figs. 18(c,d) are the $C(\epsilon)$ corresponding to the
exponential used  for the simulated $P(T)$ which in this case has $\Gamma =
130~ KeV$. The width of the lorentzian is modified by the  oscillations and
its value depends on the position and magnitude  of the peaks of $P(T)$.
Note that to obtain   $C(\epsilon)$ in MeV we have divided the time
expressed in fm/c by $\hbar c$.

The above  considerations are important since as shown in figs. 13 and 17
very often one has peaks superimposed on an exponential-like behaviour for
the delay time  probability.  These peaks are due to the small
quasi-regular regions inside the chaotic sea which give a greater
contribution to the probability.  However they cannot be easily separated
and they influence inevitably also the behaviour of the most chaotic
trajectories. In other words the interplay of  different reaction times,
due to the strong mixing between regular and irregular motion, is in
general difficult to disentangle completely in  such a complex reaction
mechanism.

In general the claim often advanced that the lorentzian distortions of
$C(\epsilon)$ are only due to finite size effect \cite{wei92} is not
always true.

\section{Realistic calculations }

The quantal approach we have used up to now is a simplified description of
heavy ion scattering.  A more realistic model should take into account: a)
a three-dimensional description of the scattering; b) the effect of other
degrees of freedom (like vibrations or nucleon transfer) by means of an
absorption term in the potential; c) the calculation of  cross sections,
angular distributions and other observables directly comparable  with
experimental  data.

The role of absorption was studied in ref. \cite{ba92a}.  Adding an
imaginary component to the interaction it was demonstrated  that when the
absorption  at the barrier is strong enough the fluctuations in the
transition  probabilities can be completely washed out. As it is discussed
in the next section,  an important feature which has been found
experimentally in the heavy ion reactions of the kind  investigated here,
is the superficial transparency of the potential.   Thus in our case  the
assumption of a  weak absorption is a very realistic  approximation.
Semiclassically this means that long lived trajectories give an appreciable
contribution.

In ref. \cite{ba92b,ba92c} it has been shown that  fluctuations in the
transition probabilities are concentrated around the barrier. Due to this
reason, increasing the initial orbital angular momentum produces a shift in
the energy range where these fluctuations are clustered. This is why when
one sums over the angular momenta  to calculate cross sections
\cite{ba92b,ba92c} and angular distributions  \cite{var91} very complicated
and irregular  structures appear  again as a function of energy. In the
following, in order to consider a very realistic  quantum description of
the reaction between a deformed nucleus and  a spherical one,  we use the
three-dimensional coupled channel code  FRESCO \cite{tho88}. The latter is
a sophisticated program which can be considered the quantal analog of the
three-dimensional  classical picture described in sections II and III. At
the same time  it gives us the possibility  to take into account the three
points discussed above.

The elastic transition probability  calculated by using  FRESCO is shown in
fig.22  for the systems $^{28}Si + {^{24}Mg}$ and $^{12}C + {^{24}Mg}$.
The total angular momentum considered is $L$=10 $\hbar$.  The real part of
the nuclear ion-ion potential is the same used for the 2D calculations.
The tail of the Coulomb interaction is taken properly into account. A small
absorption is considered taking an imaginary potential $W$ which has a
Woods-Saxon shape: $W(r)= W_0 / [1 + exp((r-r_0)/a) ]$, with $W_0$=0.2 MeV,
$r_0$=0.86 $(A_1^{1/3} + A_2^{1/3})$ fm and a=0.2 fm.  The parameters used
are close to those adopted in ref. \cite{pol84} for a similar system. We
have considered only two rotational states,  the $2^+$  (at E=1.26 MeV) and
the $4^+$ (E=4.21 MeV) in the deformed nucleus  plus the ground state
$0^+$.  Only a coupling between the nearest neighbours is considered. The
total number of exit channels is 9 \cite{ba92c}. The coupling factor is
again given by eq.(21).

In fig.22  oscillations   similar to those present in the simpler 2D
calculation of fig.18 are shown.  The energy step (in the center of mass
frame) used is  0.046 MeV and 0.053 MeV for $^{28}Si + {^{24}Mg}$ and
$^{12}C + {^{24}Mg}$ respectively.  In this case, due to the absorption
considered, we have less structures and they have a smoother behaviour than
in the 2D calculations presented here where the absorption was  neglected.
However, when a summation over the  angular momenta is performed and the
cross section at backward angles is calculated,  very complicated
fluctuations appear. Examples of excitation function are displayed in
figs.23 and 24 for the same systems.

As found in the planar case \cite{ba92c}  (not shown here for lack of
space),  cross sections fluctuate irregularly for energies greater than the
barrier in all the channels considered.  In general no qualitative changes
are found in going from 2D to 3D.  However, in the latter case, the cross
sections oscillate on a larger scale.

In order to study in a quantitative way these fluctuations,  we  divide
each point of the cross section  by its local  average value.  That is we
consider the quantity  $X(E) = {d \sigma / d \Omega \over {< d\sigma / d
\Omega >}}$. This local average should be taken over an  interval $\Delta$E
which is bigger than the average width of the structures and much smaller
than the full energy range considered. This procedure eliminates the smooth
behaviour of the cross section and  at the same time enables one to
investigate the fluctuations  of a quantity which is dimensionless and vary
over a few units \cite{gla90}. The fluctuations thus obtained are shown in
fig.25 for the system  $^{28}Si + {^{24}Mg}$. The actual local average is
done over an energy interval $\Delta E=0.8$  MeV.

One can now proceed in evaluating the autocorrelation functions.  In this
case, we adopt the standard formula used for cross sections
\cite{pap88,gla90}
\begin{equation}
    {<X(E) ~ X(E+\epsilon)>\over <X(E)> ~<X(E+\epsilon)> }-1 ~.
\end{equation}
\noindent
These  autocorrelation functions are displayed in fig.26  for the states
considered in fig.25. The dashed curves are  lorentzians whose widths  are
also reported in the figure.  The autocorrelation functions  displayed in
fig.26  are different from those  shown in fig.19. The latter go to zero
and follow more closely a lorentzian shape, even though the widths of the
lorentzians for  the inelastic channels are comparable.  However the
meaning  of $C(\epsilon)$  in the two case is deeply different. In fact the
one shown in fig.19 refers to the S-matrix  for  a fixed orbital angular
momentum  (equal to the total one since I=0)  while the other correspond to
the cross section obtained summing over many (60 in this case) $\ell$.  In
heavy--ion  scattering, due to the large size of the nuclei, a great number
of waves contribute.  Therefore any realistic calculation should take it
into account a sum over numerous $\ell$-waves.

Another quantity which can be determined experimentally is the angular
distribution, i.e the differential cross section as a function of the
detection angle for a fixed energy. By means of the code FRESCO we have
calculated elastic and inelastic  angular distributions as a function of
the incident energy.  It is found a strong oscillating behaviour as the
energy is above the barrier and for backward angles.  In general the
angular distribution at large angles is dominated by the nuclear
interaction, while the Coulomb one predominates at forward angles.  Then
the backward angles fluctuations are strictly connected to the internal
part of the interaction which classically shows a chaotic dynamics.

In fig. 27 the elastic angular distributions  as a function of incident
energy  are shown. Only the angles in the range   with $86^o  <
\theta_{cm}  <  178^o$  are plotted in order to outline the irregular
behaviour. similar features were found in ref. \cite{var91}  for the 2D
quantal model.

In conclusions, we have demonstrated that  no drastic change appear in the
qualitative features of the scattering in passing from 2D to 3D. An
irregular behaviour in cross section and angular distributions persist and
can be connected to the underlying chaotic classical scattering.  In the
next section we review the main experimental features of heavy-ion
scattering around the Coulomb barrier for nuclei of the kind considered
here.

\section{Experimental overview and discussion }

In nuclear physics cross section fluctuations have been observed since the
60s, when nucleon-nucleus reactions started to be intensively studied
\cite{erb84}. Predicted by Ericson \cite{eri60,bri63,eri63,eri66},
fluctuations in compound nucleus cross  sections  were detected
\cite{co62,cas63,bre64} at excitation energies above the neutron
evaporation  barrier. That is in the energy region of strong  overlapping
resonances, where the level spacing D is very small in comparison with the
level width $\Gamma$, $\Gamma /D  \gg  1$. Fluctuations are generated  by
the random action of the very many intermediate levels which connect  the
entrance and the exit channels. According to Ericson's theory,
autocorrelation functions of experimental data  have a lorentzian shape
whose width $\Gamma$, the coherence length,  gives the energy range within
which the intermediate levels are excited  coherently. Therefore $\Gamma$
represents the average level width of the intermediate compound nucleus and
gives information on the average lifetime of the compound  nucleus
$\tau=\hbar/ \Gamma$ and on the level density.  Fluctuations have a
statistical nature, but are experimentally  reproducible.  However,
experiments with heavier projectiles - performed almost at the same time -
revealed  excitation function fluctuations with different features. The
first system to be studied  was $^{12}C + {^{12}C}$ \cite{bro60}. In this
case  fluctuations started around the Coulomb barrier presenting structures
with widths of different sizes.  In general these structures, which were
present in several reaction  channels,  became broader as the incident
energy  increased. The coherence lengths extracted  from these experiments
were larger than those  previously found in nucleon-nucleus scattering -
100-300 KeV against  10-50 KeV-   and correlation analyses showed a
nonstatistical origin. Similar characteristics were observed for $^{12}C +
{^{16}}O$, and  $^{16}O + {^{16}O}$  among several other systems
\cite{erb84}. Due to the peripheral kind of these reactions and the unusual
strongly  attractive nucleus-nucleus potential at large distances, it was
postulated that these oscillating  structures should have a molecular-like
nature substantially  different from that of the average compound nucleus.

Going to heavier systems, a more complex behaviour was detected.  In
correspondence of  excitation function fluctuations, anomalous large and
highly  oscillating angular distributions  were observed. Typical examples
of this behaviour  are the systems $^{16}O + {^{28}Si}$ \cite{bra77} and
$^{12}C +  {^{28}Si},{^{32}S}$ \cite{ost79}, where these features were
first observed. Again a dinuclear molecule was thought to be the origin,
but the  mechanism soon appeared much more complicated: systems leading to
the same nuclear composite showed different structures; it was not always
possible to understand the angular distributions in terms of only one
single wave, on the  contrary several angular momenta around the grazing
value were involved \cite{mer81}. The  phenomenon has been intensively
studied and, as in the case of Ericson's fluctuations, a  vast literature
can be found on the subject. Fundamental review papers, both on the many
experiments performed and  the  theoretical models proposed  to explain
heavy--ion resonances, are those of Erb and Bromley \cite{erb84} and
Braun-Munzinger and Barrette  \cite{bra82}.  They say clearly that
fluctuating phenomena in light systems seem to have a common nature:  there
are only quantitative, but not qualitative  differences  from system to
system. However, notwithstanding the great effort spent during these years,
there is not yet a quantitative theoretical understanding of this
behaviour: all the  advanced models fail - partly or completely - in
reproducing the  large set of existing data.  The only model-independent
consideration which  comes out naturally from the experimental analysis is
the unexpected presence of a very weak surface absorption. In other words,
a  relatively small number of channels is involved.

Though the interest in these intriguing phenomena diminished in the 80s,
some groups have continued the experimental research. Thus fluctuations
were recently found in the  elastic and inelastic cross sections of heavier
nuclei like $^{28}Si + {^{28}Si}$ \cite{bet81}, $^{24}Mg + {^{24}Mg}$
\cite{zur83} and $^{24}Mg + {^{28}Si}$ \cite{wuo87}. At the same time
excitation function fluctuations were observed also in  deep inelastic
collisions  of several systems like $^{19}F + {^{89}Y}$ \cite{suo87},
$^{28}Si + {^{64}Ni}$, $^{28}Si + {^{48}Ti}$ \cite{pap88,car89}. Again,
differences from Ericson's theory were found,  mainly because correlations
between  several channels and a clear angular dependence were evidenced.
In ref. \cite{gla90} cross section fluctuations were measured for several
windows of energy loss, establishing this way a connection between
oscillating  phenomena in elastic and damped  reactions.

The connection between fluctuations in light heavy--ion reaction and a
chaotic mechanism though generically addressed already in ref. \cite{blu88}
was stressed for the  first time in the  conclusions of ref. \cite{gla90}.
In sections II-VI of this paper we have presented a model which exhibits
chaotic scattering and is able to reproduce in a semiquantitative way the
experimental phenomenology for light heavy--ion reaction discussed above.
The puzzling irregularities observed experimentally find a natural
explanation in the framework of chaotic scattering considering only a few
degrees of freedom.

One could think that having used  rotational states  chaotic scattering is
limited only to this kind  of excitations. Actually,   features very
similar   to those here discussed have been found both classically
\cite{das92} and quantum-mechanically \cite{das93} for heavy--ion reactions
considering vibration modes.

We can therefore conclude that irregular scattering  has a well established
theoretical and experimental  foundation in light heavy-ion collisions.

In general the single fluctuations    are not theoretically reproducible -
quantum chaos seems to maintain a strong sensitivity on the input
parameters -  and one should compare instead autocorrelation functions,
widths distributions or other statistical  quantities.  However absorption
can help in  increasing the theoretical predictive power smearing out the
wildest fluctuations.

Ericson fluctuations are often claimed to be the quantum manifestation of
classical chaotic scattering.  Actually Ericson's theory was proposed  for
compound nucleus reactions and considers only the universal statistical
aspects  of chaotic scattering moreover, as already mentioned, it applies
only for strongly overlapping resonances.  In this sense it has been
related to the Random Matrix Theory \cite{blu88,smi91,lew91}. In our
approach direct, semidirect and long lived  reactions are  taken explicitly
into account. Partially broken invariant surfaces which  correspond to what
is usually called soft chaos \cite{gut90} seem to be  present.  Moreover
the fact that in our case $\Gamma/D \le 1$ indicates  a regime of
chaoticity produced by a dynamical mechanism which differs from the
Ericson's one. We want to stress that both regular and fully chaotic
scattering are two extreme  exceptional cases. In general one finds more
often a mixed situation which lies in between. This situation is the most
complicated to deal with, expecially in the quantum case where the
chaos-to-order transition  is  more elusive.   In this respect a lot of
work has still to be done in order to characterize quantitatively this
transition.

\section{Summary}

It has been shown that chaotic scattering represents  a real possibility in
collisions between light heavy ions and that it can explain  the irregular
fluctuations observed experimentally.  A few degrees of freedom can
generate a very complicated and unpredictable  motion expecially  when
semiclassical approximations are used. This is an important result both for
nuclear physics  and for more fundamental questions like the existence and
the features of  quantum chaos. These investigations  allow to reinterpret
standard approaches - although  for the moment only in a generic way - in
the new framework of the  transition from order to chaos.  The study of
heavy--ion scattering is particularly interesting due to its  privileged
position  between the classical and the quantum world.


\begin {table}
\caption{Parameters used for the deformed nuclei studied in the text.}
\label {tab1}
\begin{tabular}{|cccc|}
  nucleus ~~~       & $\alpha_{20}$   & $Q_o$      & $\Im \hbar^{-2}$   \\
  ~              &  ~              & $(fm^2)$   & $(MeV^{-1}) $      \\
\hline
  $^{24}Mg ~~~$     & ~0.42 ~         & ~ 57.  ~   & ~2.378 ~         \\
  $^{252}Sm~~~$     & ~0.246 ~        & ~ 360. ~   & ~25.   ~         \\
\end{tabular}
\end{table}

\newpage

\begin {table}
\caption{Characteristic quantities in the case of 2D
classical scattering for
 E= 25 MeV and   L=0 $\hbar$, see text. }
\label {tab2}
\begin{tabular}{|ccccccc|}
$\Delta \Phi_i$ ~~~~ & $D^u$  & $D^s$    &  $\overline{D}$  & $\Gamma$
& $\overline{\lambda}$ & z  \\[2ex]
 (deg) ~~~~        & ~       & ~        &   ~      & $10^{-3}$ (c/fm)
& $10^{-3}$ (c/fm)& \\[2ex]
\hline
 10  ~~~~          & ~0.79 ~ & ~ 0.85  ~& ~0.82 ~  & ~0.4 ~
     &~ 2.20 ~       &    ~3.14~     \\[2ex]
 1   ~~~~          & ~0.83 ~ & ~ 0.89  ~& ~0.86 ~  &~ 0.48 ~
         & ~3.43 ~     &     ~3.33~     \\[2ex]
 0.1 ~~~~          & ~0.84 ~ & ~ 0.73  ~& ~0.79 ~  &~ 0.4 ~
       & ~ 1.88 ~     &     ~4.3 ~     \\[2ex]
 0.01 ~~~~         & ~0.91 ~ & ~ 0.77  ~& ~0.84 ~  &~ 0.37 ~
        & ~ 2.3 ~     &     ~4.34~     \\[2ex]
\end{tabular}
\end{table}

\begin {table}
\caption{Characteristic quantities in the case of
3D classical scattering for
 E= 25 MeV,  L=0 $\hbar$, $\phi_i=90^o$ and $\Phi_i=45^o$. See text. }
\label {tab3}
\begin{tabular}{|ccccccc|}
$\Delta \Phi_i$ ~~~~   & $D^u$  & $D^s$    &  $\overline{D}$ & $\Gamma$
     & $\overline{\lambda}$  & z    \\[2ex]
 (deg)                 & ~       & ~        &   ~ &  $10^{-3}$ (c/fm) &
$10^{-3}$ (c/fm) & ~ ~  \\[2ex]
\hline
 ~~~10  ~~~            & ~0.82 ~ &~ 0.83  ~& ~0.82 ~     & ~0.38 ~
      &~ 2.10 ~          & ~2.98 ~   \\[2ex]
 ~~~1   ~~~            & ~0.82 ~ &~ 0.77  ~& ~0.79 ~     &~ 0.35 ~
     & ~1.70 ~          & ~2.7 ~   \\[2ex]
 ~~~0.1 ~~~            & ~0.82 ~ &~ 0.73  ~& ~0.77 ~     &~ 0.28 ~
   & ~1.23 ~          & ~2.57 ~   \\[2ex]
 ~~~0.01 ~~~           & ~0.91 ~ & ~0.70  ~& ~0.81 ~     &~ 0.33 ~
   & ~1.71 ~          & ~3.41 ~  \\[2ex]
\end{tabular}
\end{table}

\newpage

\begin{figure}
\label{f1}
\noindent
\caption{Coordinate system used. Polar coordinates $r$, $\theta$ and $\phi$
specify the position of the spherical projectile  nucleus, while $\Theta$
and $\Phi$  are the Euler angles of the  intrisic frame of the deformed
target nucleus.}
\end{figure}

\begin{figure}
\label{f2}
\noindent
\caption{The ion-ion potential adopted is plotted for  the system  $^{28}Si
+ {^{24}Mg}$. Three value of orbital angular momentum $\ell=15,35,45
{}~\hbar$ are shown for two  orientation angles, i.e. $\xi=0^o$ (dashed
curve), and $\xi=90^o$ (full curve). }
\end{figure}

\begin{figure}
\label{f3}
\noindent
\caption{The final scattering angle $\phi_f$ is plotted as a function of
the  initial rotor orientation  $\Phi_i$ for the reaction  $^{28}Si
+{^{24}Mg}$. For a fixed total angular momentum $L=15~ \hbar$, three
different energy values are considered. See text. }
\end{figure}

\begin{figure}
\label{f4}
\noindent
\caption{The deflection function for the same system of fig.3 at E=28 MeV,
a).  Two successive magnifications are also shown in  b) and c). See text.}
\end{figure}

\begin{figure}
\label{f5}
\noindent
\caption{Poincar\'e surfaces of section for 10 trajectories bound inside
the interaction zone. The system considered is $^{28}Si + {^{24}Mg}$.
Different values of the deformation parameter $\alpha$ are used in order to
illustrate the order-to-chaos transition in the interaction region.  More
precisely $\alpha /\alpha_{20}=1,0.15,0.1$ going from top to bottom panel,
being $\alpha_{20}$ the real deformation value of  ${^{24}Mg}$ used in this
paper.  }
\end{figure}

\begin{figure}
\label{f6}
\noindent
\caption{Magnification of the surface of section shown in fig. 5 (middle
panel) for $\alpha /\alpha_{20}=0.15 $. In this case 90 trajectories are
considered.}
\end{figure}

\begin{figure}
\label{f7}
\noindent
\caption{Final values of the rotor spin $I_f$ (in units of the maximum spin
$I_{max}= {E\over {2 \Im}}$), the deflection angle  $\phi_f$ and the
reaction time $T_f$ as a function of the initial rotor orientation for the
2D scattering of the system $^{4}He +{^{24}Mg}$. The values of the total
energy and angular momentum are E=6 MeV  and L=5 $\hbar$ respectively.}
\end{figure}

\begin{figure}
\label{f8}
\noindent
\caption{ The same of fig.7 for the system $^{12}C +{^{24}Mg}$. In this
case  the total energy and the total angular momentum are  E=12 MeV and L=5
$\hbar$ respectively.}
\end{figure}

\begin{figure}
\label{f9}
\noindent
\caption{ The same of fig.7 for the system $^{86}Kr +{^{24}Mg}$. In this
case  the total energy and the total angular momentum are  E=63 MeV and
L=10 $\hbar$ respectively.}
\end{figure}

\begin{figure}
\label{f10}
\noindent
\caption{ The same of fig.7 for the system $^{86}Kr +{^{152}Sm}$. In this
case  the total energy and the total angular momentum are   E=270 MeV and
L=0 $\hbar$ respectively.}
\end{figure}

\begin{figure}
\label{f11}
\noindent
\caption{Final values of the rotor spin, the deflection  angles ($\phi_f$
and $\theta_f$) and the reaction time versus the initial rotor orientation
for the reaction  $^{12}C +{^{24}Mg}$ considering 3D scattering. The total
energy is E=5 MeV and  the total angular momentum is   L=12 $\hbar$.}
\end{figure}

\begin{figure}
\label{f12}
\noindent
\caption{The final scattering angle $\phi_f$ is shown  as a function of the
initial conditions for various small intervals.  The reaction is $^{28}Si
+{^{24}Mg}$ at E=25 MeV  and zero total angular momentum. In (a-d) and
(e-h) 2D and 3D is considered. In the 3D case the initial orientation
angle is $\Phi_i=45^o$. }
\end{figure}

\begin{figure}
\label{f13}
\noindent
\caption{ In the upper part (a) the classical cross section is shown as a
function of $\phi_f$,  for the case illustrated in fig.12(c). In the lower
part (b) we  plot the quantity $-ln<1/N(R)>$ versus $lnN(R)$ to calculate
the fractal dimension D by means of the sandbox method. In this case one
obtains D=0.73, see text. }
\end{figure}

\begin{figure}
\label{f14}
\noindent
\caption{  The behaviour of $Log_{10}(f/\epsilon)$ versus
$Log_{10}(\epsilon)$ is shown for the same case displayed in fig.13.  The
slope,  which in this case is D=0.84, gives the uncertainty dimension,  see
text.}
\end{figure}

\begin{figure}
\label{f15}
\noindent
\caption{The reaction time probability distributions for the intervals of
initial conditions shown in fig.12. The straight lines are linear fits
whose slope gives the escape rate $\Gamma$ reported in tables 2 and 3.}
\end{figure}

\begin{figure}
\label{f16}
\noindent
\caption{The reaction time probability distribution for the case shown in
fig.12(f) and 15(f) is displayed in comparison with two different fits:  an
exponential law (a) and a power law (b) are adopted.  }
\end{figure}

\begin{figure}
\label{f17}
\noindent
\caption{  The function $\lambda(T)$ is shown versus time for  ten
trajectories  randomly chosen in the intervals of initial conditions
considered in fig.12(d) and fig.12(h)  in order to estimate the Lyapunov
exponents.  Both in the 2D (a) and in the 3D (b) case an  asymptotic value
$\lambda_{\infty}$ is approached.  The value  $\overline{\lambda}_{\infty}$
shown represents the average over the  asymptotic limits of the
trajectories considered.  }
\end{figure}

\begin{figure}
\label{f18}
\noindent
\caption{  Quantal elastic transition probability as a function  of
incident energy.  The calculations are the result of the 2D coupled
channels approach described in the text. An energy step equal to 20 KeV is
used.  }
\end{figure}

\begin{figure}
\label{f19}
\noindent
\caption{Autocorrelation functions (open squares) corresponding  to the
transition  probabilities shown in the previous figure. The full curves are
lorentzian fits whose width is also reported. See text. }
\end{figure}

\begin{figure}
\label{f20}
\noindent
\caption{Classical delay time probability distributions for  the reaction
$^{28}Si +{^{24}Mg}$. The scattering is in 2D for L=15 $\hbar$  and
E=28,28.5,29  MeV (a,b,c). The respective autocorrelation functions are
plotted in  (d,e,f) as open squares, while lorentzian fits are drawn as
full curves. The widths are also reported.}
\end{figure}

\begin{figure}
\label{f21}
\noindent
\caption{Simulated delay time probability distribution (a,b) and
respective autocorrelation functions (c,d). The full curves in (c,d)
correspond to the analytical $C(\epsilon)$, while the open squares refer to
the numerical evaluation. The dashed curves are lorentzian whose width
$\Gamma$=130 KeV  corresponds to the exponential law used in (a,b), see
text.}
\end{figure}

\begin{figure}
\label{f22}
\noindent
\caption{ Elastic transition probabilities calculated by means of  the 3D
code FRESCO (see text) for the systems $^{28}Si +{^{24}Mg}$ (a) and $^{12}C
+{^{24}Mg}$ (b).  The total angular momentum is L=10 $\hbar$.  See text. }
\end{figure}

\begin{figure}
\label{f23}
\noindent
\caption{ Excitation functions for the system $^{28}Si +{^{24}Mg}$ obtained
by means of the 3D code FRESCO, plotted for $\theta_{cm}=178^o$. The
channels $0^+$, $2^+$ and $4^+$ are displayed. See text for more details.}
\end{figure}

\begin{figure}
\label{f24}
\noindent
\caption{The same of fig.23 for the  system $^{12}C +{^{24}Mg}$ and
$\theta_{cm}=180^o$. }
\end{figure}

\begin{figure}
\label{f25}
\noindent
\caption{ The fluctuations found for the excitation functions  displayed in
fig.23 are evidenced by dividing the cross section $d\sigma/d\Omega$  by
its average local value obtained by considering an  interval $\Delta E$=0.8
MeV. See text.}
\end{figure}

\begin{figure}
\label{f26}
\noindent
\caption{Autocorrelation functions corresponding to the fluctuations
displayed in fig.25 (full squares). Lorentzian curves (dashed)  are shown
for comparison. The widths are also reported. }
\end{figure}

\begin{figure}
\label{f27}
\noindent
\caption{Elastic angular distributions, with $86^o < \theta_{cm} < 178^o $,
as a funtion of the  incident energy. The calculations refer to the
reaction $^{28}Si +{^{24}Mg}$  and were done by means of the 3D code
FRESCO.  }
\end{figure}


\begin{references}
\bibitem{lyapnote} In the 3D case, the fact that  $\overline{\lambda} =
\overline \Gamma /(1-\overline{D}) \le \overline{\lambda}_{\infty} $ can be
explained if more than one expanding direction exists. Indeed in this case
$\overline \Gamma =\sum_i \overline \lambda_i/ (1-\overline D_i)$  and
taking only the first term,  as we did, the inequality  $\overline \Gamma
\le  \overline\lambda_{\infty} (1- \overline D_i) $  should hold (T. T\'el
{\it private communication}).
\end{references}
\end{document}